\documentclass[aps,prl,showpacs,twocolumn,floatfix]{revtex4}

\usepackage{amsmath}
\usepackage{bm}
\usepackage{mathrsfs}
\usepackage{graphicx}
\usepackage{subfigure}

\renewcommand{\bar}[1]{\overline{#1}}
\providecommand{\Journal}[4] {#1 {\bf#2}, #3 (#4)}
\providecommand{\PLB}{Phys. Lett. B} %
\providecommand{\PRL}{Phys. Rev. Lett.} %
\providecommand{\PRD}{Phys. Rev. D}
\providecommand{\EPJC}{Eur. Phys. J. C} %
\providecommand{\ZPC}{Z. Phys. C} %
\providecommand{\JHEP}{J. High Energy Phys. } %
\providecommand{\JPG}{J. Phys. G} %

\begin{document}

\title{Influence of Heavy Quark Recombination on the Nucleon
Strangeness Asymmetry}
\author{Puze Gao}
\affiliation{Institute of Theoretical Physics, Chinese Academy of
Sciences, Beijing 100080, China}
\author{Bo-Qiang Ma\footnote{Email address: \texttt{mabq@phy.pku.edu.cn}}}
\affiliation{School of Physics and State Key Laboratory of Nuclear
Physics and Technology, Peking University, Beijing 100871, China}
\begin{abstract}
The nucleon strange and anti-strange distribution asymmetry is an
important issue in the study of the nucleon structure. In this work,
we show that the heavy quark recombination processes from a
perturbative QCD picture can give a sizable influence on the
measurement of the nucleon strangeness asymmetry from
charged-current charm production processes, such as the CCFR and
NuTeV dimuon measurements. When the influence of heavy quark
recombination is considered, a positive effective $\delta
S^-_{\mathrm{HR}}$ should be added to the initially extracted
strangeness asymmetry $S^-\equiv\int dx[s(x)-\bar s(x)]x$,
supporting the strangeness asymmetry $S^-$ being positive, which is
helpful to explain the NuTeV anomaly within the framework of the
standard model.
\end{abstract}

\maketitle

The nucleon strange and anti-strange distributions are important
quantities in the study of the nucleon structure, and a clear
knowledge of them helps for a better understanding of some related
phenomena in experiments. The nucleon strange quark-antiquark
distribution asymmetry is predicted naturally by some
non-perturbative models~\cite{bm96}, and a positive strangeness
asymmetry $S^-\equiv\int dx[s(x)-\bar s(x)]x$ has been
shown~\cite{kre04,oln03,dm04,alw04,dxm04,wak04} to be a promising
mechanism to explain the NuTeV anomaly~\cite{NT1,NT2} within the
framework of the standard model. Perturbative QCD at three-loops can
also generate a strangeness asymmetry~\cite{pQCD04}, however, the
obtained magnitude is one order smaller to be relevant to the NuTeV
anomaly.

In the measurement of the nucleon strangeness asymmetry, some
valuable works have been done, though no conclusive result has been
reached. Since strangeness asymmetry should be a very small quantity
in inclusive deep inelastic scattering (DIS) cross sections, it is
difficult to be extracted precisely. However, (anti-)neutrino
induced charged current charm production processes are quite
sensitive to the (anti-)strange distribution, and thus can provide
valuable information on the strangeness asymmetry. CCFR and NuTeV
dimuon measurements~\cite{CCFR95,CCFR93,NTdimu01} are of such
experiments. Although earlier analysis of dimuon data did not show
support of the nucleon strangeness
asymmetry~\cite{CCFR95,NTdimu01,mason04}, a recent next to leading
order (NLO) analysis of the NuTeV data with improved method does
show some evidence of the nucleon strangeness
asymmetry~\cite{mason06}
$S^-=0.00196\pm0.00046({\mathrm{stat}})\pm0.00045({\mathrm{syst}})\pm0.00128({\mathrm{external}})$.
Meanwhile, global analysis have indicated positive strangeness
asymmetry~\cite{bar00,oln03,lai07}, such as the most recent work of
Lai \textit{et al.}~\cite{lai07}, who include both CCFR and NuTeV
dimuon data sets in their analysis, and produce the allowed range of
$-0.001<S^-<0.005$ at 90\% confidence level. These analysis suggest
that $S^-$ is likely positive.

In this work, we aim at checking the measurement of the nucleon
strangeness asymmetry by including a perturbatively calculable QCD
effect. We find that the heavy quark recombination processes can
produce a sizeable influence on the measurements with
charged-current charm production process such as the CCFR and NuTeV
dimuon measurements.

Heavy quark recombination~\cite{br02b,br02c,br02prl} combines a
heavy quark, e.g., c quark, with a light antiquark $\bar q$ (or
$\bar c$ with q) of relative small momentum in the hard scattering,
and the $(c\bar q)$ subsequently hadronizes into a D meson.
Refs.~\cite{br02c,br02prl} employ simple perturbative QCD pictures
and explain the charm photoproduction  asymmetry and the leading
particle effect~\cite{E791,gm07} successfully. In the following, we
show how the heavy quark recombination influences the measurement of
the nucleon strangeness asymmetry.

The CCFR and NuTeV dimuon measurements have provided important
information on the strangeness degrees of freedom in the parton
structure of the nucleon. These measurements both rely on the
(anti-)neutrino induced charged-current charm production processes,
with the leading order (LO) subprocesses being $\nu_\mu +s(d)
\rightarrow \mu^- +c$ and $\bar{\nu}_\mu +\bar{s}({\bar d})
\rightarrow \mu^+ +\bar{c}$. The produced c($\bar c$) quark
hadronizes and then decays partially into $\mu^+(\mu^-)$ to form a
second $\mu$. The oppositely signed dimuon events in experiment are
then recorded for analysis of the nucleon strange distributions.

The CCFR and NuTeV experiments use iron as their target, and for
simplicity, we take it as an isoscalar target. The strange quark
antiquark distribution asymmetry is directly related to the
difference between neutrino and anti-neutrino induced dimuon
differential cross section at LO, which can be expressed
as~\cite{gm05}
\begin{eqnarray}\label{eq:SALO}
{d^2\sigma_{\nu_{\mu}N\rightarrow \mu^- \mu^+ X}\over d\xi dy }-
{d^2\sigma_{\bar\nu_{\mu}N\rightarrow \mu^+ \mu^- X}\over d\xi dy }
 ={G_F^2 S\over \pi r_w^2}f_cB_c
\nonumber
\\ \times
\left\{ \xi[s(\xi)-\bar s(\xi)]|V_{cs}|^2+{1\over
2}\xi[d_v(\xi)+u_v(\xi)]|V_{cd}|^2\right\},
\end{eqnarray}
where $\xi$ is the light-cone momentum fraction of the struck quark
and is related to the Bjorken scaling variable $x$ through $\xi=
x(1+m_c^2/Q^2)$, $S=2ME_\nu$ and $y=\nu/E_\nu$ with $E_\nu$ and
$\nu$ being  the incident energy and the energy transfer in the
nucleon rest frame, $r_w\equiv 1+Q^2/M_W^2$ and $f_c\equiv 1-{m_c^2/
2ME_\nu\xi}$, and $B_c$ is the branching ratio for $c\rightarrow
\mu^+ X$. The valence contribution in Eq.~(\ref{eq:SALO}) is
suppressed relative to strange contribution from their relative
coefficients $|V_{cs}|^2\sim 0.9$ and $|V_{cd}|^2\sim 0.05$, thus
this cross section difference of Eq.~(\ref{eq:SALO}) is sensitive to
the strange distribution asymmetry $S^-(\xi)\equiv \xi[s(\xi)-\bar
s(\xi)]$.

The heavy quark recombination processes as
\begin{equation}\label{eq:nuHR}
\nu_\mu+\bar q\rightarrow \mu^-+\bar s(\bar d)+D(c\bar q),
\end{equation}
\begin{equation}\label{eq:nubarHR}
\bar\nu_\mu+ q\rightarrow \mu^++ s(d)+\bar D(\bar c q) ,
\end{equation}
(diagrams in FIG. 1) can also contribute to the dimuon final states
through $D(\bar D)$ decays: $D\rightarrow\mu^+X$ and $\bar
D\rightarrow\mu^-X$. These processes are possible to have sizable
effect in the extraction of the nucleon strangeness asymmetry.
\begin{figure}
  {\includegraphics[width=.65\columnwidth]{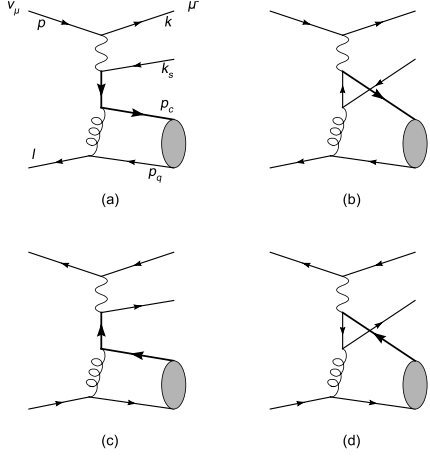}}\hspace{15mm}
  \caption{(a) and (b) are diagrams for $c\bar q $ recombination into a D meson in neutrino induced process~(\ref{eq:nuHR});
  (c) and (d) are diagrams for $\bar c q$ recombination into a $\bar D$ meson in anti-neutrino induced process~(\ref{eq:nubarHR}).
  Thick lines are heavy quarks, and shaded blobs are D or $\bar D$ mesons.}
\end{figure}

The processes of (\ref{eq:nuHR}) and (\ref{eq:nubarHR}) contribute
to the difference between  neutrino and anti-neutrino induced dimuon
differential cross section at higher order, which can be expressed
as
\begin{eqnarray}\label{eq:csdHR}
&&\left [{d^2\sigma_{\nu_{\mu}N\rightarrow \mu^- \mu^+ X}\over d\xi
dy }- {d^2\sigma_{\bar\nu_{\mu}N\rightarrow \mu^+ \mu^- X}\over d\xi
dy}\right]_{HR}\nonumber\\
&=&\sum_{q,D}\int dx[\bar q(x)-q(x)]{d^2\hat{\sigma}_{D(c\bar
q)}\over d\xi dy} B_{D(c\bar q)},
\end{eqnarray}
where $d\hat{\sigma}_{D(c\bar q)}$ denotes the cross section for the
subprocess (\ref{eq:nuHR}), which is identical to the cross section
of subprocess (\ref{eq:nubarHR}) from charge symmetry. $B_{D(c\bar
q)}$ is the branching ration for $D(c\bar q)\rightarrow \mu^+X$. q
denotes a light quark flavor from the nucleon, which could be u or
d, and the $D(c\bar q)$ meson could be either a scalar $^1S_0$ state
or a vector $^3S_1$ state.

Such a contribution as Eq.~(\ref{eq:csdHR}) serves as an additional
part in the extracted strange distribution asymmetry, such as the
NLO analysis of the NuTuV dimuon data~\cite{mason06}, because the
recombination processes as FIG. 1 are not included in the ananysis.
Thus the realistic strange distribution asymmetry
$S^-_{\mathrm{real}}(\xi)$ should be the analysed result
$S^-_{\mathrm{analy}}(\xi)$ minus the contribution from heavy quark
recombination processes, which is a negative quantity from
Eq.~(\ref{eq:csdHR}) since the nucleon structure ensures $\bar
q(x)-q(x)<0$ for $q=u,d$. From Eq.~(\ref{eq:SALO}) and
Eq.~(\ref{eq:csdHR}), one gets
\begin{equation}\label{eq:SArl}
S^-_\mathrm{real}(\xi)=S^-_\mathrm{analy}(\xi)+\delta
S^-_\mathrm{HR}(\xi),
\end{equation}
with
\begin{eqnarray}\label{eq:SAHR}
\delta S^-_\mathrm{HR}(\xi)&\approx &
 {\pi r_w^2\over G_F^2 S f_c B_c|V_{cs}|^2}\nonumber\\
&\times&\sum_{q,D}\int dx[q(x)-\bar q(x)]{d^2\hat{\sigma}_{D(c\bar
q)}\over d\xi dy} \cdot B_{D(c\bar q)},~~~~
\end{eqnarray}
where $\delta S^-_\mathrm{HR}(\xi)>0$ since to minus a negative
quantity is equivalent to plus a positive quantity. Thus the
realistic strangeness asymmetry $S^-_\mathrm{real}\equiv\int d\xi
S^-_{\mathrm{real}}(\xi)$ should be larger than the experimentally
extracted value according to the contribution from the unaccounted
recombination processes.

Now we proceed to estimate the size of $\delta
S^-_\mathrm{HR}(\xi)$. We follow the method in Ref.~\cite{br02b} to
calculate the heavy quark recombination process. For the color
singlet $^1S_0$ $D(c\bar q)$ production, the following substitution
is made in the parton amplitude:
\begin{equation}\label{eq:substitution}
v_j(p_q)\bar u_i(p_c)\rightarrow x_q{\delta_{ij}\over N_c}m_c
f_+(p\!\!\!/_c-m_c)\gamma_5.
\end{equation}
Then set $p_q=x_qp_c$ in the amplitude and take the limit
$x_q\rightarrow 0$. Thus the amplitude for color singlet $^1S_0$
state $D(c\bar q)$ production is (diagrams in FIG. 1(a)(b)):
\begin{eqnarray}\label{eq:amplitue}
M_{in}&=&{16\pi G_F\alpha_s m_c\delta_{in}f_{+}\over
9\sqrt{2}r_w(2l\cdot p_c)}
L^\mu \bar v(l)\gamma^\nu(p\!\!\!/_c-m_c)\gamma_5 \nonumber\\
&&\times [\gamma_\nu
{p\!\!\!/-k\!\!\!/-k\!\!\!/_s+m_c\over(p-k-k_s)^2-m_c^2}\gamma_\mu(1-\gamma_5)\nonumber\\
&& +\gamma_\mu(1-\gamma_5){l\!\!\!/-k\!\!\!/_s\over
(l-k_s)^2}\gamma_\nu ]v(k_s),
\end{eqnarray}
where $L^\mu=\bar u(k)\gamma^\mu(1-\gamma_5)u(p)$ is the lepton
current.

The $\delta_{ij}$ in Eq.~(\ref{eq:substitution}) is the color
factor for color-singlet state, which is replaced by
$\sqrt{6}T_{ij}^a$ for color-octet state together with the
nonperturbative parameter $f_+$ replaced by $f_+^8$.
$\rho_1=f_+^2$ and $\rho_8=(f_+^8)^2$ will appear in cross
sections to characterize the probability for a color-singlet and a
color-octet $^1S_0(c\bar q)$ state to hadronize into a state
including a $^1S_0$ state $D(c\bar q)$ meson. The subprocess cross
section for $^1S_0$ state $D(c\bar q)$ meson production thus can
be expressed as
\begin{equation}\label{eq:factor18}
d\hat{\sigma}_{D(c\bar q)}=d\hat{\sigma}[c\bar q(^1S_0)_1]\cdot
\rho_1+d\hat{\sigma}[c\bar q(^1S_0)_8]\cdot \rho_8.
\end{equation}
The $d\hat{\sigma}[c\bar q(^1S_0)_8]$ can be calculated to be
different from $d\hat{\sigma}[c\bar q(^1S_0)_1]$ by a single color
factor of 1/8. Thus, $d\hat{\sigma}_{D(c\bar q)}$ of
Eq.~(\ref{eq:factor18}) can be expressed as
\begin{equation}\label{eq:factoreff}
d\hat{\sigma}_{D(c\bar q)}=d\hat{\sigma}[c\bar q(^1S_0)_1]\cdot
\rho_{\mathrm{eff}}[c\bar q(^1S_0)\rightarrow D(c\bar q)],
\end{equation}
with $\rho_{\mathrm{eff}}=\rho_1+\rho_8/8$.

For $^3S_1$ state production of vector meson $D^*(c\bar q)$, similar
substitution as Eq.~(\ref{eq:substitution}) with the $\gamma_5$
replaced by $\epsilon\!\!\!/$ is made in the parton amplitude, where
$\epsilon$ is the polarization vector for the $^3S_1$ state. Similar
expression as Eq.~(\ref{eq:factoreff}) can be obtained for the
subprocess cross section of vector $D^*(c\bar q)$ meson production,
\begin{equation}
d\hat{\sigma}_{D^*(c\bar q)}=d\hat{\sigma}[c\bar q(^3S_1)_1]\cdot
\rho_{\mathrm{eff}}[c\bar q(^3S_1)\rightarrow D^*(c\bar q)].
\end{equation}

Physically, there may be spin-flipped transitions such as $c\bar
q(^1S_0)\rightarrow D^*(c\bar q)$. While we have neglected such
transitions, partly because the calculation of charm
photoproduction~\cite{br02c} and the leading particle
effect~\cite{br02prl} have both set $\rho_{\mathrm{sf}}=0$, and
partly because the inclusion of these transitions will not greatly
affect our result, since both $D(c\bar q)$ and $D^*(c\bar q)$ meson
will decay similarly to $\mu^+$.

The flavor-changing transitions, such as $c\bar u\rightarrow
D^+(c\bar d)$, are also neglected as Refs.~\cite{br02c,br02prl},
because these transitions are relatively suppressed in the large
$N_c$ limit of QCD, and also because the inclusion of such
transitions will not affect our result notedly.

The number of free parameters can be greatly reduced from symmetries
of the strong interaction. As discussed in Ref.~\cite{br02c}, heavy
quark spin symmetry implies
\begin{equation}
\rho_{\mathrm{eff}}[c\bar q(^1S_0)\rightarrow D(c\bar
q)]=\rho_{\mathrm{eff}}[c\bar q(^3S_1)\rightarrow D^*(c\bar q)],
\end{equation}
and SU(3) light quark flavor symmetry indicates, for example,
\begin{equation}
\rho_{\mathrm{eff}}[c\bar u(^1S_0)\rightarrow
D^0]=\rho_{\mathrm{eff}}[c\bar d(^1S_0)\rightarrow D^+].
\end{equation}
Thus, only one parameter is left:
\begin{eqnarray}
\!\!\!\rho_{\mathrm{sm}}&\equiv&\rho_{\mathrm{eff}}[c\bar
d(^1S_0)\rightarrow
D^+]=\rho_{\mathrm{eff}}[c\bar d(^3S_1)\rightarrow D^{*+}]\nonumber\\
&=&\rho_{\mathrm{eff}}[c\bar u(^1S_0)\rightarrow
D^0]=\rho_{\mathrm{eff}}[c\bar u(^3S_1)\rightarrow D^{*0}]. ~~~~
\end{eqnarray}
Thus, for isoscalar target, the $\delta S^-_\mathrm{HR}(\xi)$ of
Eq.~(\ref{eq:SAHR}) can be expressed as
\begin{eqnarray}\label{eq:SAHRex}
\delta S^-_\mathrm{HR}(\xi)\approx {\pi r_w^2\over
G_F^2 S f_c |V_{cs}|^2 B_c}\int dx[u_v(x)+d_v(x)]\nonumber\\
\times\left[{d\hat{\sigma}[c\bar q(^1S_0)_1]\over d\xi
dy}b_1+{d\hat{\sigma}[c\bar q(^3S_1)_1]\over d\xi dy}b_2\right]\cdot
\rho_\mathrm{sm},
\end{eqnarray}
where $b_1={(B_{D^+}+B_{D^0})/2}$ and
$b_2={(B_{D^{*+}}+B_{D^{*0}})/2}$.

The subprocess cross section can be calculated straight forward
from the parton amplitudes (with the parameter $f_+$ extracted out
for $\rho_\mathrm{sm}$). Since the subprocess is a $2\rightarrow
3$ process, there are five independent variables in the subprocess
cross section. From the symmetry of the scattered $\mu^-$ around
the incident direction, four independent variables are left, where
two variables are transformed to $\xi$ and $y$ (or $Q^2$) and the
other two are integrated out. Thus for $\delta
S^-_\mathrm{HR}(\xi)$ of fixed $Q^2$, the integration is totaly of
3 dimensions including the integral on $x$ in
Eq.(\ref{eq:SAHRex}). The boundaries of the integration are
determined from the allowed physical phase space.

We use the CTEQ6L parton distributions for the nucleon~\cite{CTEQ6},
and the running coupling constant $\alpha_s$ is as specified in
CTEQ6L. We take $m_c=1.5$ GeV and set the factorization scale to be
$\sqrt{p_{c\perp}^2+m_c^2}$. Since the two muons in NuTeV experiment
are required to have energy greater than 5 GeV, we try similar cuts
for the produced $\mu$ and the charmed meson in our integration. We
find that the cross section from heavy quark recombination process
decreases very slowly with the increase of the cut on the energy of
the produced charmed meson. Thus the recombination processes are not
suppressed by the cuts in experiments.

FIG. 2 shows our result of $\delta S^-_{\mathrm{HR}}(\xi)$ for
$E_\nu=160$ GeV, $Q^2=20$ GeV and $\rho_{\mathrm{sm}}=0.15$. Such
$E_\nu$ and $Q^2$ are approximate averaged incident energy
 and $Q^2$ in the NuTeV dimuon
experiment~\cite{NTdimu01}. $\rho_{\mathrm{sm}}$ is the
nonpertubative parameter for the heavy quark recombination and
$\rho_{\mathrm{sm}}=0.15$ is the LO fitted result from charm
photoproduction asymmetry~\cite{br02c}. The branching ratios and
$|V_{cs}|$ are taken to be the central values from
Ref.~\cite{PDG06}.
\begin{figure}
{\includegraphics[width=0.80\columnwidth]{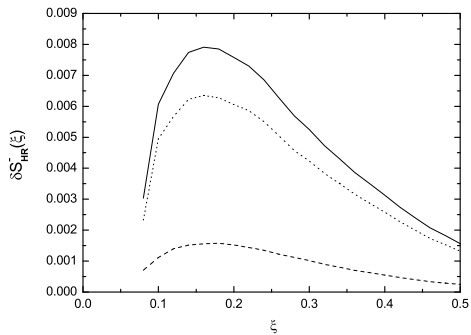}}
 \caption{$\delta S^-_{\mathrm{HR}}(\xi)$ for
$E_\nu=160$ GeV, $Q^2=20$ GeV$^2$ and $\rho_{\mathrm{sm}}=0.15$. The
dashed curve is the contribution from $^1S_0$ state; the dotted
curve is the contribution from $^3S_1$ state; and the solid curve is
their sum, the $\delta S^-_{\mathrm{HR}}(\xi)$.}
\end{figure}

From FIG. 2, one sees that $\delta S^-_{\mathrm{HR}}(\xi)$ is a
valence-like distribution, with its peak in the range of
$\xi=0.1\sim0.3$. From Eq.~(\ref{eq:SArl}), the realistic strange
distribution asymmetry is the sum of the analysed result of dimuon
experiments and the effective contribution from heavy quark
recombination $\delta S^-_{\mathrm{HR}}(\xi)$. We can estimate
$\delta S^-_{\mathrm{HR}}$ by integrating $\delta
S^-_{\mathrm{HR}}(\xi)$ over $\xi$, and get $\delta
S^-_{\mathrm{HR}}\approx 0.0023$ for $\rho_{\mathrm{sm}}=0.15$.

Such a value of $\delta S^-_{\mathrm{HR}}$ significantly enhances
the measured strangeness asymmetry to a larger positive value, since
$S^-_{\mathrm{real}} =S^-_{\mathrm{analy}}+\delta
S^-_{\mathrm{HR}}$. Recent NLO analysis of the NuTeV dimuon data
provides positive strangeness asymmetry centered at
0.00196~\cite{mason06}. With the correction of the heavy quark
recombination, the central value of the realistic strangeness
asymmetry could be $S^-_\mathrm{real}\approx0.0043$.

Such a value of the strangeness asymmetry can explain the NuTeV
anomaly to a large extent. NuTeV anomaly arises from the large
discrepancy of the NuTeV measurement of the $\sin\theta_w$ with the
standard model prediction, and becomes a hot debated area in recent
years. The NuTeV measurement of the $\sin\theta_w$ relies on the
hypothesis that strange and anti-strange distributions are
symmetric. When this assumption is violated, their result on
$\sin\theta_w$ will change. The influence of nonzero $S^-(\xi)$ to
the result of $\sin\theta_w$ is most sensitive in the range $\xi=
0.06-0.3$ (FIG. 1 in Ref.~\cite{NT2}), and such range is just the
position of the peak of $\delta S^-_{\mathrm{HR}}(\xi)$. Thus the
non-vanishing $\delta S^-_{\mathrm{HR}}(\xi)$ and
$S^-_\mathrm{real}$ may have a large effect in the NuTeV measurement
of $\sin\theta_w$. In such sensitive range, a positive strangeness
asymmetry $S^-$ of the order 0.005 can fill the gap between theory
and the experiment of the NuTeV anomaly. The value of $\delta
S^-_{\mathrm{HR}}\approx 0.0023$ for $\rho_{\mathrm{sm}}=0.15$ alone
can provide nearly half of the strangeness asymmetry needed to
explain the NuTeV anomaly.

In our calculation, there are some uncertainties in the choice of
the factorization scale $\mu$ and the parameter $m_c$. The impact
for different choice of $m_c$ on our result is small (within 5
percent for $m_c=1.5\pm0.3$~GeV). While for different choice of the
factorization scale, the influence is large. Ref.~\cite{br02c}
calculate charm photoproduction with factorization scale
$\sqrt{p_{\perp}^2+m_c^2}$, where $p_{\perp}$ is the transverse
momentum of the produced D relative to the incident photon
direction. In the process of this work, the photon is replaced by
the W boson, and thus the calculation we take is performed for
$\mu_0= \sqrt{p_{c\perp}^2+m_c^2}$, where $p_{c\perp}$ is the
transverse momentum of the produced D relative to the W boson
direction in the nucleon rest frame. The result nearly trebles when
$\mu=\mu_0/2$, and the result reduces nearly by half when
$\mu=2\mu_0$. The uncertainties may imply that higher order effects
are still important. Ref.~\cite{br02c} reports small effect on the
predicted asymmetry when varying the factorization scale, where only
the ratios of the cross sections are concerned, and scale dependence
might be canceled in that case.

The parameter $\rho_{\mathrm{sm}}$ still has some uncertainty in its
value. In FIG. 2, we use $\rho_{\mathrm{sm}}=0.15$, which is from
the fit of charm photoproduction asymmetry at LO~\cite{br02c}. As
discussed in Ref.~\cite{br02c}, there is at least 30\% uncertainty
in this parameter $\rho_{\mathrm{sm}}$ due to finite heavy quark
mass, SU(3) breaking and $1/N_c$ corrections, and more over,
$\rho_{\mathrm{sm}}$ should be multiplied by a $K$ factor if NLO
corrections in photo-gluon fusion are incorporated in their
calculation. Thus, parameter $\rho_{\mathrm{sm}}$ could well be as
large as 0.3. Such a value for $\rho_{\mathrm{sm}}$ means $\delta
S^-_{\mathrm{HR}}\approx 0.0046$, which alone is sufficient to
explain the NuTeV anomaly. On the other hand, $\rho_{\mathrm{sm}}$
could be smaller than 0.15, such as the LO fit from the leading
particle effect~\cite{br02prl}, $\rho_1=0.06$, where $\rho_1$ is the
parameter for color singlet state, which is different from
$\rho_{\mathrm{sm}}$ but their size  should be compatible. If we
take $\rho_{\mathrm{sm}}=0.06$, we get $\delta
S^-_{\mathrm{HR}}\approx 0.0009$. Such a $\delta S^-_{\mathrm{HR}}$
alone is too small to explain the NuTeV anomaly, however, this
$\delta S^-_{\mathrm{HR}}$ could still shift the dimuon result to a
larger positive value, and make the allowed range of $S^-$ entirely
positive. More precision determination of the parameters for the
heavy quark recombination and a reanalysis of the dimuon events with
consideration of the heavy quark recombination processes will be
helpful to a better knowledge of the nucleon strangeness asymmetry.

Our work implies the significance of using heavy quark recombination
mechanism~\cite{br02b,br02c,br02prl}, i.e., a perturbatively
calculable QCD effect, to reveal the strangeness asymmetry. Let us
recall that a previous LO analysis~\cite{NT2} of nucleon strangeness
asymmetry by NuTeV collaboration reported a negative value
$S^{-}=-0.0027 \pm 0.0013$, whereas their new NLO
analysis~\cite{mason06} gave a positive value centered at 0.00196 as
mentioned above. The difference between the two values is 0.0047,
which is of the same order as the $\delta S^-_{\mathrm{HR}}$
estimated in this work. This again supports our work to take higher
order effects into account.

In summary, we investigated the influence of heavy quark
recombination in (anti-)neutrino induced charged current charm
production processes on the measurement of the nucleon strange
distribution asymmetry. Our result shows that the influence could be
quite sizable and the realistic strangeness asymmetry $S^-$ should
be larger than the initially experimental results. From our
investigation and the result of recent experimental analysis, the
nucleon strangeness asymmetry $S^-$ should be positive and could be
large enough to explain the NuTeV anomaly.

We thank Y.-Q. Chen, Y.-J. Gao and Y. Jia for helpful discussions.
This work is partially supported by National Natural Science
Foundation of China (Nos.~10421503, 10575003, 10528510), by the Key
Grant Project of Chinese Ministry of Education (No.~305001), and by
the Research Fund for the Doctoral Program of Higher Education
(China).

\end{document}